# How the Quantum Sorites Phenomenon Strengthens Bell's Argument

Malcolm Forster

University of Wisconsin-Madison

February 28, 2014

**Abstract**:  A recent theorem by Colbeck and Renner (2011) leads to a significantly stronger theorem than Bell's famous theorem; the new theorem does not assume Outcome Independence.  It is suggested that the reason that the stronger theorem was not discovered earlier is that it exploits a very strong prediction of quantum mechanics by applying some mathematical "tricks" to bypass the need to use Outcome Independence.  The very strong prediction of quantum mechanics is aptly described as the "quantum Sorites" phenomenon.

**1. Introduction**:  Recently, Colbeck and Renner (2011) published a theorem that strengthens the Bell (1964) theorem in a way that is more significant than the other variations of Bell's theorem that have been published in the 50 years since Bell's celebrated theorem.  Colbeck and Renner themselves do not relate their theorem directly to Bell's theorem at all, so here I present a version of the Colbeck-Renner theorem that makes the relationship clear and explicit.  I will refer to this version of the Colbeck-Renner theorem as the *Stronger Theorem*.

For those already familiar with Bell's argument, the significance of the Stronger Theorem is summarized as follows:  It derives a contradiction without using the assumption of Outcome Independence, also known as Completeness (Jarrett (1984)), or Causality (van Fraassen (1982)).  This means that the contradiction cannot be resolved by *only* rejecting Outcome Independence (O.I.), at least not if one thinks that hidden variables can improve the predictions of quantum mechanics (QM). This is highly significant because rejecting O.I. is often seen as the most plausible way of maintaining the peaceful coexistence of quantum mechanics and relativity (see, for example, Jarrett (1984), Butterfield (1989), and more recently, Ghirardi (2010)).

Rather than presenting the Bell argument first, and then showing how it can be strengthened, I will do the opposite.  I present the stronger argument first, partly because it is simpler in many ways, and partly because it puts what's new in the paper up front.  Then I will explain how Bell's result follows from the Stronger Theorem, and explain how and why the Stronger Theorem is stronger.

The Stronger Theorem exploits the strength of QM predictions, in particular the prediction of perfect correlations.  Perfect correlations are also predicted in standard Bell experiments.   Graßhoff, Portmann and Wüthrich (2005) have shown that this leads to a relatively minor strengthening of Bell's original argument.  Their argument is expanded and generalized in Hofer-Szabó, Rédei and Szabó (2013).  It may appear at first sight that these arguments also do away with the assumption of O.I. But this appearance is wrong because their theorems are famed in terms of "common causes" and O.I. is built in as a part of the way they define "common causes".

Perfect correlations are also predicted in the GHZ version of the Bell experiment (Greenberger, Horne and Zeilinger (1989)), but the phenomenon exhibited in this experiment is *still* not strong enough to allow the Stronger Theorem to apply (see the Appendix).

I argue that the Stronger Theorem applies only in the context of the *quantum Sorites phenomenon*, otherwise known as the chained, or iterated, Bell phenomena. (The name "quantum Sorites" is newly coined here, but the idea goes back in the literature at least as far as (Braunstein and Caves (1990)). The main purpose of this paper is to *explain* why the quantum Sorites phenomenon is very different.

**2. The Quantum Sorites Phenomenon**. The Bell argument is a reductio ad absurdum argument that derives a contradiction from the main assumption that the probabilistic predictions of quantum mechanics (QM) are correct *in a particular set of experiments*. The contradiction follows from this main premise only after three premises about the properties of a hypothetically postulated hidden variable, $\lambda$, are added. The important difference between the Stronger Theorem and Bell's Theorem is that the probabilistic predictions of QM refer to a new set of experiments in which QM predicts an extreme kind of phenomenon, which I am calling the quantum Sorites phenomenon. It is the strength of the quantum Sorites phenomenon that makes it possible to derive a contradiction from significantly weaker assumptions about the properties of the variable $\lambda$. So, there is a sense in which the *total* set of premises have the same *total* strength in both theorems. It is the way the total strength is divided up between the assumed QM predictions and the assumptions about $\lambda$ that makes the difference. Since the QM predictions are so well confirmed, nobody (or almost nobody) wants to remove the contradiction by denying that the QM predictions are true, so the arrow of modus tollens must be directed against the theoretical assumptions involving $\lambda$. The weaker these assumptions are, the stronger the conclusion of the argument (that at least one of the assumptions involving $\lambda$ is false).

In particular, it is the strength of the predicted Sorites *probabilities* that allows to the premises involving $\lambda$ to be substantially weaker. As we shall see in section 4, there is a provable sense in which it is impossible for the quantum Sorites probabilities to be strengthened (and, interestingly, the same cannot be said about the probabilities in the GHZ version of the Bell experiment; see the Appendix).

The quantum Sorites phenomenon arises for correlated electron pairs in the singlet state, and for correlated pairs of photons. I will describe it for correlated photon pairs. The basic experimental setup is shown in Fig. 1. Two photons are created at a source situated between two measurement devices; one photon travels towards each of those devices. The device on the left is in Alice's laboratory, and she has control over the setting for her device. The setting is adjusted by moving a lever on the side of device—it merely adjusts the angle of a polaroid filter, or polarizer, inside the device. The angles may vary from anything from 0° to 90°; this angle is the value of the variable *A*. Similarly, Bob, on the right, has control over the angle of his polarizer, which also ranges over angles from 0° to 90°. Neither experimenter has (direct) control over the outcome of their measurement, which is recorded by the value of the variable *X* for Alice, and *Y* for Bob. 0 is recorded if the photon is blocked by the polarizer, and 1 when it passes through the analyzer. The outcome is made visible by a light on the device flashing 'red' when the photon is blocked and 'green' when the photon passes through. We are assuming that every pair of photons produces an outcome. Of course, the actual experimental design is more complicated than the one described here—my aim is to make the essential features as comprehensible as possible to non-physicists.

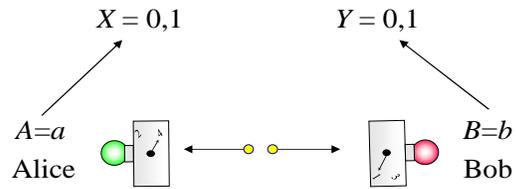

**Figure 1**: Two photons fly out towards measurement devices, one controlled by Alice on the left, and the other controlled by Bob on the right.

To understand the quantum Sorites phenomenon, one only needs to know two elementary facts about these correlated photon pairs. (1) When the angle between Alice's and Bob's measurement directions (the directions along which their polarizers are oriented) is sufficiently small, the outcomes of their respective measurements are very probably the same, $P(X = Y) \approx 1$. I will say more about the meaning of "sufficiently small" and "very probably correlated" later. (2) The outcomes are perfectly anti-correlated, $P(X \neq Y) \approx 1$, when the difference in the angles is 90°.

The quantum Sorites phenomenon arises from a *chained* set of the measurement settings like the one schematically represented in Fig. 2. Since $N\Delta\theta = 90°$, if $\Delta\theta = 2°$, then $N = 45$. In that case, there are 46 measurement angles in total, 23 for Alice and 23 for Bob, and 46 chained experiments in total. They are chained in the sense that each angle for Bob is linked to two adjacent angles for Alice, and similarly for Alice's angles, as represented by the solid lines in Fig. 2, and finally the two ends of the chain ($A = 90°$ and $B = 0°$) are linked together (the dashed line). All of these experiments except the last will (very probably) produce a matched pair of outcomes ($X = Y = 0$ or $X = Y = 1$), while the last one (the dashed line in Fig. 2) will always produce a mismatch ($X \neq Y = 0$ or $X \neq Y = 1$).

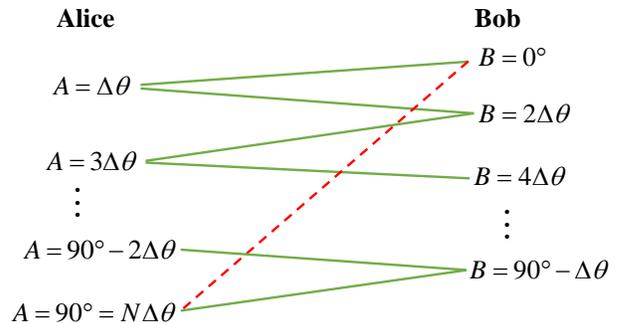

**Figure 2**: The quantum Sorites phenomenon arises from a chained set of experiments.

The probabilities of detecting a difference Alice and Bob's outcomes when the difference in the measurement angles is small is proportional to the square of the difference of the angles, $\Delta\theta$. The exact value of the QM probability of observing a mismatch is $\sin^2(\Delta\theta)$, which is approximately equal to $(\Delta\theta)^2$ when $\Delta\theta$ is small. Thus, the probability function, shown in Fig. 3, is very flat for small angles. For small angles, it approximates the Simplified QM curve also shown in Fig. 3. The Simplified QM probability function has a striking property. It says that the probability of a match between the outcomes is exactly 1 in every experiment except the last one, in which case the probability is 0.

Because the QM probability is equal to $(\Delta\theta)^2$, which is proportional to $1/N^2$, QM exhibits the same properties as Simplified QM in the limit as *N* tends to infinity (the supplement for Colbeck and Renner (2011) has details of the proof). In the limit, the probability that *none* of the *N* correlations is broken tends 1 as *N* tends to infinity.

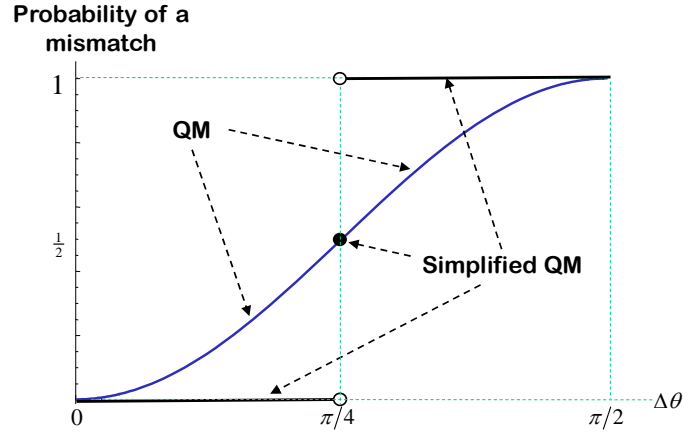

**Figure 3**: The probability of a mismatch between Alice's and Bob's outcome as a function of the difference in their measurement angles.

The fact that pairs of photons can "orchestrate" their behavior in this way is truly remarkable given that the measurement events going on in Alice's laboratory may be space-like separated from the measurement events going on in Bob's laboratory. The analogy in the next section is designed to explain why this phenomena implies some kind of non-local action at a distance.

**3. The Caffè Latte Example**. Consider an ESP experiment in which Alice and Bob are taken into separate rooms, and asked if they want a latte. They can answer 'yes' or 'no'. This will be recorded as the outcome of the experiment. Alice wants to answer the same as Bob if the latte she is offered is very close to the strength of Bob's latte ('yes' iff he says 'yes'). Otherwise, she wants to give the opposite answer ('no' iff he says 'yes'). Most of the time, Alice will be offered a latte that is almost the strength of Bob's. The single exception is when Alice is offered a 3-shot latte and Bob is offered a 0-shot latte. Suppose they decide beforehand that Bob will accept his latte no matter what. If Alice is offered anything less than a 3-shot latte, she will accept it because it will be close to what Bob is offered, and she knows he will accept it. But, what can she do when she's offered the 3-shot latte? If Bob is offered something a little less than a 3-shot latte, she wants to say 'yes'. But if Bob is offered a 0-shot latte, she wants to say 'no'. She needs to know what Bob is offered. So, she asks the experimenter what Bob was offered? The experimenter doesn't know. So she guesses. She guesses right. Repeatedly, she gets it right. All cases in which she say 'yes' are cases in which Bob is offered something a little less than a 3-shot latte, and all the cases in which she says 'no' are cases in which Bob is offered a 0-shot latte. How is this possible? What's the explanation? We seem forced to say that she somehow tacitly knows what Bob is offered via some kind of ESP (if every other form of communication is ruled out).

As a prelude to what is to come, note that it does not help Alice if she is told that Bob accepted his latte (Bob's outcome). She needs to know the measurement setting for Bob (what kind of latte he is offered) in order to decide. In other words, it does not matter whether Outcome Independence is violated or not. If Alice is able to reliably choose correctly, then there must be some *other* kind of nonlocality involved.

Nor is the puzzle resolved by allowing Alice and Bob to plan differently when they are in the waiting room. They might decide that Bob will say 'yes' when his latte is less than 2 shots and 'no' otherwise, for example. Alice is still faced with the same dilemma when she is offered a 3-shot latte. Or, it might be Alice that always says 'yes' unconditionally, but then Bob faces the dilemma whenever he is offered a 0-shot latte. And so on. No matter what they do, they will face the same kind of problem.

The behavior of correlated photons is puzzling for exactly the same reason. The theorem in Bell (1964) is essentially a proof that there is no way around this problem.

The quantum Sorites phenomenon provides a particularly vivid illustration of the Bell argument for quantum nonlocality. However, it is not merely that the quantum Sorites phenomenon provides a *vivid* illustration of Bell's argument. It leads to a *logically stronger* version of the argument!

**4. The Colbeck-Renner Lemma**. What I am calling the Colbeck-Renner Lemma is a simple theorem of probability theory. It is the mathematical trick that will allow the Stronger Theorem to bypass the assumption of Outcome Independence (O.I.) used in the standard Bell theorems.

*Lemma*: Assume that *X* and *Y* are two-valued variables.

(a) If $P(X = Y) = 1$, then $P(X = 1) = P(Y = 1)$.

(b) If $P(X \neq Y) = 1$, then $P(X = 1) = 1 - P(Y = 1)$.

The proof is omitted. The important fact is that the result follows solely from the axioms of probability; no other assumptions are used. The important feature of the Lemma is that it takes us from a fact about the joint probability distribution of *X* and *Y* to a relationship between the marginal distributions of *X* and of *Y* separately. Also note that the Lemma applies to conditional probabilities, provided that all the probabilities are conditional on the same proposition and the probabilities are well defined.

Here is an illustration of how the Lemma works. For all the settings in the chain of experiments represented by a solid line in Fig. 2, $P(X = Y | A = a, B = b) = 1$ in the limit in which the difference in the measurement angles tends to 0. From the Lemma, part (a), it follows that

$$P(X = 1 | A = a, B = b) = P(Y = 1 | A = a, B = b).$$

This is an experimentally well confirmed fact. There is no mention of any hidden variable $\lambda$, so I shall follow van Fraassen (1982) in referring to these probabilities as *surface* probabilities. Note that there is no constraint placed on the values of these marginal probabilities. They can be anything from 0 to 1. The only constraint is that they are equal to each other.

We have assumed that these conditional probabilities are well defined. Let us call this assumption Weak Surface Autonomy (for reasons that will be clear later).

*Weak Surface Autonomy*: Let the set $S$ ($S$ stands for 'Settings') be the set of angle pairs $(a,b)$ that define the chain of experiments in Fig. 2. Then, for all $(a,b) \in S$, $P(A=a, B=b) > 0$.

There is another assumption about the surface probabilities that is also experimentally very well confirmed. It is commonly called the no-signaling condition, but following van Fraassen (1982), I shall call it Surface Locality.

*Surface Locality*: For all $(a,b) \in S$,

$$P(X=1 | A=a, B=b) = P(X=1 | A=a) \text{ and } P(Y=1 | A=a, B=b) = P(Y=1 | B=b).$$

From Surface Locality, we know, for example, that the probability of Bob's outcome given that his polarizer is set at $B = 0°$ is the same in both experiments in which he uses that setting. That is,

$$P(Y=1 | A=\Delta\theta, B=0°) = P(Y=1 | A=90°, B=0°) = P(Y=1 | B=0°) = p_0$$

Similarly, for Alice,

$$P(X=1 | A=\Delta\theta, B=0°) = P(X=1 | A=\Delta\theta, B=2\Delta\theta) = P(Y=1 | A=\Delta\theta) = q_1.$$

Here $p$ codes for 'Bob' and $q$ codes for 'Alice' and the subscript refers to their measurement angle in multiples of $\Delta\theta$. Now, a remarkable conclusion follows from what I call the Sorites Theorem.

*The Sorites Theorem*: If $p_0 = q_1 = p_2 = \cdots = p_{N-1} = q_N = 1 - p_0$ then $p_0 = \frac{1}{2} = q_1 = p_2 = \cdots = q_n$.

Again, the proof is omitted, but it is important to note that this result also follows from no special assumptions. So, what Weak Surface Autonomy and Surface Locality prove is that *all* the marginal probabilities much be ½. The proof depends on two things. The first is the fact that the surface correlations are perfect. The second is that the experiments are chained together in the way depicted in Fig. 2.

If these conditions are met, then there is a sense in which the surface probabilities are *maximally* strong. To explain this, first note that perfect correlations are maximal since these involve probabilities of 1, and such probabilities are the highest you can have. On the other hand, the marginal probabilities are ½, and these are maximally *weak* in the sense that any probability different from ½ half would count as stronger. However, the Sorites Theorem tells us that there is a necessary tradeoff between the strength of the probabilities of the correlations and the strength of the marginal probabilities; namely, if the correlation probabilities are maximal, then the marginal probabilities must be ½. We don't have the option of strengthening the marginal probabilities while maintaining the strength of the correlations.

Of course, if the marginal probabilities were 1 or 0, and the probabilities of the correlations were also 0 or 1, then those probabilities would be stronger in an absolute sense. That is what everyone hoped deterministic hidden variable theories would achieve. The quantum Sorites phenomenon explains that, while this absolute level of strength is not achieved in QM, there is nevertheless a conditional sense in which the predictions of QM are maximally strong.

**5. The Stronger Theorem**. In the quantum Sorites phenomenon, for the experiments where the difference in the measurement angles is sufficiently small, we have perfect correlations,

$$P(X = Y \mid A = a, B = b) = 1.$$

Now consider *any* logical partition of propositions $\Lambda$, and denote an arbitrary member of $\Lambda$ by $\lambda$. By probability theory, we may expand $P(X = Y \mid A = a, B = b)$ as an average over probabilities conditionalized on $\lambda$'s. Since the average has to be 1, the conditional probabilities that are averaged over have to be 1 (an average can never be greater than all the things averaged over, and no probability is ever greater than 1). Therefore, if $P(\lambda \mid A = a, B = b) > 0$, then $P(X = Y \mid A = a, B = b, \lambda) = 1$. Notice that Weak Surface Autonomy already tells us that $P(A = a, B = b) > 0$, for all $(a,b)$ in the given set, so the assumption that $P(\lambda \mid A = a, B = b) > 0$ is equivalent to $P(A = a, B = b, \lambda) > 0$. This looks like Weak Surface Autonomy except that it includes the hidden variable $\lambda$, so I will refer to the new assumption as Weak *Hidden* Autonomy (Weak H.A.). This assumption needs to be carefully formulated so that it plays the role that we need it to play later on.

*Weak H.A.*: Let the set $S$ be the set of angle pairs $(a,b)$ that define the chain of experiments. For all $\lambda \in \Lambda$, if $P(A = a, B = b, \lambda) > 0$ for some $(a,b) \in S$ then $P(A = a, B = b, \lambda) > 0$ for all $(a,b) \in S$.

To show that it is possible to derive a contradiction from strictly weaker assumptions than Bell's theorem, we need to say exactly what the assumptions are. The main premise is that the probabilities predicted by QM in the experiment are correct, and we have already shown that these surface probabilities are maximally strong in the sense that we cannot shift the marginal probabilities from ½ without destroying the perfect correlations.

Now consider any $\lambda \in \Lambda$ such that, for all $(a,b) \in S$,

$$P(X = Y \mid A = a, B = b, \lambda) = 1.$$

The Colbeck-Renner Lemma immediately implies that

$$P(X=1|A=a, B=b, \lambda) = P(Y=1|A=a, B=b, \lambda).$$

We can apply the Sorites Theorem if we can link these probabilities into a chain using an assumption like Surface Locality, except that in this case Surface Locality will not work because the probabilities are now conditional on $\lambda$. We need some assumption that it would be natural for us to call Hidden Locality, but it turns out that the assumption already has an established name in the literature. It is called Parameter Independence (P.I.).

*Parameter Independence* (P.I.): $P(X=1|A=a, B=b, \lambda) = P(X=1|A=a, \lambda)$ and

$$P(Y=1|A=a, B=b, \lambda) = P(Y=1|B=b, \lambda).$$

It says that, in the context in which $\lambda$ is fixed, the setting on the other wing of the experiment is probabilistically irrelevant to the probability of the outcome on this wing. This is a new assumption; it is not implied by Surface Locality.

P.I. is also a premise in Bell's theorem. But unlike in Bell's theorem, we only need to add one more very weak assumption to derive a contradiction. I call this the Improved Predictions assumption.

*Improved Predictions*: It is possible to *improve* the predictions of QM, not by showing them to be false, but by recovering the QM probabilities by averaging over probabilities conditional on $\lambda$ *in a non-trivial way* (that is, where at least one of the probabilities conditional on $\lambda$ is different from the corresponding surface probability).

Most, if not all, hidden variable theorists want to accept Improved Predictions. The existence of variables that do not improve the predictions of QM is trivial. As an example, take $\lambda$ to be the color of Bob's tie, or whether Alice was wearing her lab coat during the experiment.

Recall from section 4 that we proved that all the marginal probabilities (of a single outcome) have to be ½. Analogously, in this section, using Weak H.A. and P.I., we can obtain a similar chain of equations that, by the Sorites Theorem, imply that all the marginal probabilities conditional on $\lambda$ are ½. But this contradicts the Improved Predictions assumption, because ½ does not improve the predictions of QM. That is how we can derive a contradiction from the three assumptions Weak H.A., P.I., and Improved Predictions.

Here are the details. Using P.I., we derive consequences such as

$$q_1^\lambda = P(X=1|A=1, \lambda) = P(Y=1|B=0, \lambda) = p_0^\lambda.$$

That is $p_0^\lambda = q_1^\lambda$. From the next experiment in the chain, we infer $q_1^\lambda = p_2^\lambda$, and so on, until we get $p_{N-1}^\lambda = q_N^\lambda$. Finally, when we link the two ends together, we get $q_N^\lambda = 1 - p_0^\lambda$. So, by the Sorites Theorem, all of these probabilities is ½. Therefore, none of the probabilities given $\lambda$ improve the predictions of QM.

**6. Why the Stronger Theorem is Stronger.** The Stronger Theorem derives a contradiction from four assumptions. The first is that the QM predicted probabilities are correct. The second is Weak H.A., the third is P.I., and the fourth is Improved Predictions. Bell's Theorem also derives a contradiction from four assumptions. The first is that the QM predicted surface probabilities are correct. The second is called Hidden Autonomy (H.A.), the third is P.I., and the fourth is Outcome Independence (O.I.). P.I. has already been stated. Here is H.A. and O.I.

*H.A.*: For all $\lambda \in \Lambda$, $(a,b) \in S$, $P(\lambda | A = a, B = b) = P(\lambda)$.

*O.I.*: For all $(a,b) \in S$, and all $\lambda \in \Lambda$ such that $P(A = a, B = b, \lambda) > 0$,

$$P(X = 1, Y = 1 | A = a, B = b, \lambda) = P(X = 1 | A = a, B = b, \lambda) P(Y = 1 | A = a, B = b, \lambda).$$

The logical relationship between the assumptions of Bell's theorem and the assumptions of the Stronger Theorem is one of strict entailment. More specifically,

    (a) H.A. $\Rightarrow$ Weak H.A.
    (b) O.I. $\Rightarrow$ Improved Predictions.

Given these entailments, Bell's theorem is a simple logical consequence of the Stronger Theorem.

*Proof of* (a): Suppose $P(\lambda | A = a, B = b) = P(\lambda)$ and $P(A = a, B = b, \lambda) > 0$ for *some* $(a,b) \in S$. Then $P(\lambda) > 0$, and therefore $P(A = a, B = b, \lambda) > 0$ for *all* $(a,b) \in S$.

*Proof of* (b): Suppose that $P(A = a, B = b, \lambda) > 0$ and

$$P(X = 1, Y = 1 | A = a, B = b, \lambda) = P(X = 1 | A = a, B = b, \lambda) P(Y = 1 | A = a, B = b, \lambda).$$

It is a well-known fact about the surface probabilities that outcomes are unconditionally correlated. That is,

$$P(X = 1, Y = 1 | A = a, B = b) \neq P(X = 1 | A = a, B = b) P(Y = 1 | A = a, B = b).$$

Therefore, for at least some values of $\lambda$, the probabilities conditional are $\lambda$ must be different from the corresponding surface probabilities. Therefore, Improved Predictions holds.

    Finally, note that the entailments (a) and (b) do not work in the other direction. So, the Stronger Theorem is *strictly* stronger than Bell's theorem.

**7. Discussion.** Colbeck and Renner (2011) do not present their theorem in this form. Nor did they make its relationship to Bell's theorem clear. Ghirardi and Romano (2013) criticize Colbeck and Renner's description of the assumptions of their theorem, but do not talk about the proof, nor do they attempt to relate the Colbeck-Renner Theorem to the Bell Theorem in any rigorous way. I have attempted to address these issues by formulating a version of the theorem that is explicitly and precisely related to Bell's Theorem, and by explaining how the proof works.

The startling conclusion is that Outcome Independence (O.I.) is not needed as an assumption in the Stronger Theorem. Therefore, it is not possible to remove the contradiction by only rejecting O.I, at least not if one wants to maintain Improved Predictions (the existence of $\lambda$ that violate Improved Predictions is trivial and uncontroversial.) Therefore, hidden variable theorists who maintain Improved Predictions must reject either Weak H.A., or P.I. This means that hidden variable theorists may need to rethink the claim that rejecting only O.I. is the best way for hidden variable theories to maintain the peaceful coexistence of QM and Einstein's relativity theory (e.g., Jarrett (1984), Butterfield (1989)). As recently as 2010, page 1393, Ghirardi (2010) says that it turns out to be possible to exhibit genuinely relativistic nonlocal theories which violate only O.I. If these theories do not improve the predictions of QM (violating Improved Predictions), then this claim is consistent with the Stronger Theorem. But if they satisfy Improved Predictions, that is, they do improve the predictions of QM, then they cannot violate *only* O.I.

**Appendix**. In the GHZ experiment (Greenberger et al. (1989)), three particles (electrons) are created at the same location, and then fly apart towards 3 widely separated "measuring" devices, *A*, *B*, and *C*, each of which can be put at one of 2 settings, labeled 1 and 2. Each "measurement" device has a light bulb attached, which flashes red or green when a particle passes through it. The experimenter controls the setting at position 1 or position 2 (in the experiment, the setting is just the direction in which a magnet is orientated). Which color the bulb flashes is not controlled by the experimenter. This is the outcome of the experiment. As before, for *A*, we use the variable *X* to record the outcome, which is now $-1$ or $+1$. Likewise, for *B*, we use the variable *Y* to record the outcome, $-1$ or $+1$. Finally, for *C*, we use the variable *Z* to record the outcome, $-1$ or $+1$. Now the outcomes for the three settings 1-1-2, 1-2-1, and 2-1-1 are perfectly correlated in the following sense.

$$1 = P(Y = X \times Z \mid A = 1, B = 1, C = 2) = P(Y = X \times Z \mid A = 1, B = 2, C = 1) = P(Y = X \times Z \mid A = 2, B = 1, C = 1)$$

For the last measurement setting, 2-2-2, we get a perfect *negative* correlation.

$$1 = P(Y \neq X \times Z \mid A = 2, B = 2, C = 2).$$

The two surface assumptions, Weak Surface Autonomy and Surface Locality, also apply, but a Sorites chain is not present, so the marginal probabilities are not constrained to be ½. Of course, the experimentally confirmed quantum probabilities are ½, and that is the value predicted by QM, but that may well be because a Sorites chain of experiments *could* be constructed. The point is only that the four measurement settings in the GHZ setup, described above, are not sufficient to form a Sorites chain. This is proven by assigning one or more of the marginal probabilities a probability different from ½, and then proving that the probabilities of the correlations need not change. I only need one such example. The example I choose is to assign

$$P(X = 1 | A = 1) = 1 = P(X = 1 | A = 2).$$

To maintain the same maximal correlation probabilities, we require that

$$1 = P(Y = Z | B = 1, C = 2) = P(Y = Z | B = 1, C = 1) = P(Y = Z | B = 2, C = 1) \text{ and}$$
$$1 = P(Y \neq Z | B = 2, C = 2).$$

Now the Colbeck-Renner Lemma does apply, giving the constraints

$$P(Z = 1 | C = 2) = P(Y = 1 | B = 1) = P(Z = 1 | C = 1) = P(Y = 1 | B = 2) = 1 - P(Z = 1 | C = 2).$$

But, it is possible to satisfy these constraints (by putting each of these marginal probabilities equal to ½). Therefore, there is no inconsistency involved in putting the two marginal probabilities for $X = 1$ equal to 1. It is true that after we do that, we have hit a "local maximum". So, while there is a limit to how far we can strengthen the probabilities, the surface probabilities in the GHZ experiment (the ones predicted by QM) are not maximal. This proves that there is no Sorites chain to be discovered there, which also implies that there is no Sorites chain if we conditionalize on some hidden variable $\lambda$. Thus, we cannot contradict the Improved Predictions assumption just from Weak H.A. and P.I. Of course, if we strengthen Improved Predictions to O.I. and strengthen Weak H.A. to H.A., then we can derive a contradiction. That is just the GHZ version of Bell's Theorem.